# Variable-range-hopping conduction processes in oxygen deficient polycrystalline ZnO films

Yung-Lung Huang

*Department of Electrophysics, National Chiao Tung University, Hsinchu 30010, Taiwan*

Shao-Pin Chiu

*Institute of Physics, National Chiao Tung University, Hsinchu 30010, Taiwan*

Zhi-Xin Zhu and Zhi-Qing Li

*Tianjin Key Laboratory of Low Dimensional Materials Physics and Preparation Technology, Faculty of Science, Tianjin University, Tianjin 300072, China*

Juhn-Jong Lin

*Institute of Physics, National Chiao Tung University, Hsinchu 30010, Taiwan and*

*Department of Electrophysics, National Chiao Tung University, Hsinchu 30010, Taiwan*

**Abstract**

We have fabricated oxygen deficient polycrystalline ZnO films by the rf sputtering deposition method. To systematically investigate the charge transport mechanisms in these samples, the electrical resistivities have been measured over a wide range of temperature from 300 K down to liquid-helium temperatures. We found that below about 100 K, the variable-range-hopping (VRH) conduction processes govern the charge transport properties. In particular, the Mott VRH conduction process dominates at higher temperatures, while crossing over to the Efros-Shklovskii (ES) VRH conduction process at lower temperatures. The crossover occurred at temperatures as high as a few tens degrees Kelvin. Moreover, the temperature behavior of resistivity over the entire VRH conduction regime from the Mott-type to the ES-type process can be well described by a universal scaling law.

**I. INTRODUCTION**

Zinc oxide (ZnO) is a wide band gap semiconductor with a direct gap of ~ 3.4 eV at room temperature, and is natively *n*-type. All forms (bulks, films and nanoscale structures) of this material have recently been extensively studied, both theoretically and experimentally.[1-4] These studies are largely motivated by the potential applications of this class of materials in the emerging optoelectronic, spintronic and nanoelectronic devices. Apart from the optical and magnetic properties, it is of crucial importance to understand, and hence to tailor, the charge transport processes in both the pure (undoped) and doped ZnO materials. In this regard, the overall temperature behavior from room temperature down to liquid-helium temperatures of the electrical resistivity, $\rho(T)$, can reveal the underlying charge transport mechanisms in these materials. Previously, the temperature behavior and electrical conduction processes in ZnO films have been studied by some groups.[5-8] Thermal-activation-type conduction at temperatures around and not too far below room temperature as well as the Mott variable-range-hopping (VRH) conduction[9] at lower temperatures have been reported. To the best of our knowledge, there has been no any report on the observation of the Efros-Shklovskii (ES) VRH conduction process[10-12] at even lower temperatures where the many-body electron-electron interaction effects are expected to play a role.

Compared with $In_2O_3$ and $SnO_2$ films, the resistivities of as-grown ZnO films are more sensitive to the oxygen partial pressure applied during the deposition process, because Zn is chemically more active in an oxygen atmosphere than either In or Sn. Hence, it is important to systematically investigate the charge transport properties in a series of ZnO films deposited in various oxygen partial pressures. (It is know that the oxygen contents, together with other factors, determine the carrier concentrations, and thus the resistivities, in ZnO samples.[1,4]) In this work, we have measured $\rho(T)$ in a series of oxygen deficient, polycrystalline ZnO films over a wide range of temperature from 300 K down to liquid-helium temperatures. Our films were made to span a wide range of resistivity, with $\rho(300 \text{ K})$ varying by a factor of ~

3000 from our least resistive to our most resistive samples. Covering a wide range of sample resistivity, accompanied with a wide range of measurement temperature, allows us to gain systematic and instructive information about the charge conduction mechanisms in this class of materials. In particular, we observe a crossover from the Mott VRH conduction to the ES VRH conduction processes in our samples at temperatures as high as a few tens degrees Kelvin. Our results concerning these two types of VRH conduction processes and their smooth crossover feature are discussed below.

**II. EXPERIMENTAL METHOD**

Zinc oxide films were fabricated onto glass substrates by the standard rf sputtering deposition method. A ZnO target (99.99% purity) was used as the sputtering source. The base pressure of the vacuum chamber was better than $1\times10^{-4}$ Pa before the sputtering process was initiated, and the temperature of the substrates was held at 550 °C during deposition. To obtain ZnO films with different oxygen contents (and thus, different electron donor concentrations), the films were deposited in a mixture of argon and oxygen (99.999%) atmosphere. To ensure a fine control of the oxygen flux, two currents of gases were transmitted into the gas mixing chamber: one is a pure Ar gas, another is a mixture gas of $O_2$ and Ar with a volume ratio of 2 : 98. The total flux (i.e., the pure Ar gas plus the mixture gas) was kept at 50 standard-state cubic centimeter per minute (SCCM). The pressure in the vacuum chamber was maintained at 0.55 Pa, and the volume ratios of $O_2$ to Ar were tuned to a few fixed values (0, 0.04%, 0.2%, 0.3%, 0.5%, 1.0%, and 1.6%, corresponding to an oxygen flux from the $O_2$–Ar mixture gas being set at 0, 0.02, 0.10, 0.15, 0.25, 0.50, and 0.80 SCCM, respectively) between 0 and 1.6% during the deposition processes. The thicknesses of the films were determined by a surface profiler (Dektak, 6 m), and are listed in Table I. X-ray diffraction (XRD) measurements were made using a powder diffractometer (D/MAX-2500X) with Cu $K\alpha$ radiation at room temperature.

Four-probe dc resistance measurements were carried out by utilizing a Keithley K-220 or K-6430 as a current source and a high-impedance (TΩ) Keithley K-617 or K-6430 as a voltmeter. The current leads and voltage leads were attached to the films (typically, ~ 1 cm long and ~ 3 mm wide) with silver paste. The films were placed on a sample holder which was situated inside a dark vacuum can. The vacuum can was mounted on a standard $^4$He cryostat. The temperature was monitored with a calibrated Si diode. It should be noted that the resistances reported in this work were all measured by scanning the current-voltage (*I-V*) curves at various fixed temperatures. The resistance at a given temperature was then determined from the regime around the zero bias voltage, where the *I-V* curve was linear. (In fact, our *I-V* curves in every sample were linear over a wide range of bias voltage). Since the resistances of those ZnO films deposited at high $O_2$ pressures were too large to be accurately measured down to very low temperatures, the lowest measurement temperatures for the samples No. 5, No. 6 and No. 7 were 5, 10 and 20 K, respectively.

**III. RESULTS AND DISSCUSSION**

Figure 1 shows the XRD patterns for two representative films (No. 4 and No. 7). Clearly, only the x-ray diffractions corresponding to the (002) and (004) planes of the hexagonal wurtzite structure are observed, indicating that our films are single phased and possess a good c-axis texture.

Figure 2 shows the temperature dependence of resistivity for seven ZnO films deposited at different oxygen partial pressures (see Table I). This figure clearly reveals that the resistivities increase monotonically with decreasing temperature over the whole measurement temperature range, suggesting typical semiconducting behavior in all samples. In the film No. 1 which was deposited in a pure argon atmosphere, the resistivity increases only by a factor of 2.5 between 300 K and 2 K. On the other hand, in the film No. 6 which was deposited in an 1% $O_2$ atmosphere, the resistivity increases by about four orders of magnitude as the sample is cooled from 300 K down to 10 K. Overall, at a given fixed measurement temperature, the resistivities of the films increase with increasing oxygen contents, as expected. In the following discussion, we shall concentrate on temperatures below about 100 K in order to focus on the Mott and the ES VRH conduction processes which we observed in the samples Nos. 3–7.

Generally speaking, in a semiconductor in the extrinsic regime and at sufficiently low temperatures, the charge transport between localized states in an impurity band arises owing to a few charge carriers which hop from some occupied states below the Fermi level, $E_F$, to some unoccupied states above the $E_F$. The nearest-neighbor-hopping (NNH) process may be the dominant conduction mechanism at relatively higher temperatures.[13] When the temperature is reduced, the number of empty sites among nearest neighbors for a given occupied state could become too few (and also the thermal phonon energies are not sufficiently high), leading to a freezing of the NNH conduction. Then, the VRH conduction processes may take place and play a key role in the charge transport. In three-dimensional systems (which are pertinent to our samples) and in the absence of the many-body electron-electron interaction effects, the temperature dependence of the VRH conduction (which is a phonon-assisted quantum-mechanical tunneling process) predicated by Mott is given by[9]

$$\rho_M(T) = \rho_{M_0} e^{(T_M/T)^{1/4}}, \quad (1)$$

where $\rho_{M_0}$ is a resistivity parameter, and $T_M$ is a characteristic temperature. Figure 3 shows the logarithm of resistivity as a function of $(1/T)^{1/4}$ for our five most resistive ZnO films (the samples Nos. 3–7). It is clearly seen that the resistivities obey the $\log \rho \propto (1/T)^{1/4}$, or $\ln \rho \propto (1/T)^{1/4}$, law between ~ 30 and ~ 90 K. The experimental data for these samples were least-squares fitted to Eq. (1) (the straight solid lines), and we found that the minimum temperatures above which the Eq. (1) remains valid are approximately 20, 28, 32, 40 and 40

K for the samples No. 3, No. 4, No. 5, No. 6 and No. 7, respectively. Below these minimum temperatures, our experimental data deviate from the predications of Eq. (1), as can be readily seen in Fig. 3. That is, the Mott VRH conduction law can no longer describe the charge transport properties in these ZnO films at low temperatures. Our fitted values for the relevant parameters are listed in Table I. We note in passing that, for our least resistive samples No. 1 and No. 2, the resistivities can not be described by Eq. (1) in any temperature interval (see below for further discussion).

In the Mott VRH conduction theory, the characteristic temperature is defined by[9,13,14]

$$T_\mathrm{M} = \frac{18}{k_B N_0(E_F)\xi^3}, \qquad (2)$$

where $N_0(E_F)$ is the electronic density of states (DOS) at the Fermi level and in the absence of electron-electron interactions, and $\xi$ is the localization length of the relevant electronic wavefunction. The most probable hopping distance and the average hopping energy are, respectively, given by[9,13,14]

$$\overline{R}_{hop,Mott} = \frac{3}{8}\xi\left(\frac{T_M}{T}\right)^{1/4} \approx \frac{3}{8}a_B\left(\frac{T_M}{T}\right)^{1/4}, \qquad (3)$$

and

$$\overline{W}_{hop,Mott} = \frac{1}{4}k_B T\left(\frac{T_M}{T}\right)^{1/4}, \qquad (4)$$

where we have taken the electronic localization length to be the effective Bohr radius, $a_B$, of the shallow donors. In ZnO, $a_B \approx 2$ nm for the major shallow donors, such as oxygen vacancies, Zn interstitials, and hydrogen impurity atoms.[15] Using the measured values of $T_M$, our experimental values of $N_0(E_F)$, $\overline{R}_{hop,Mott}$ and $\overline{W}_{hop,Mott}$ are calculated and listed in Table I. Inspection of Table I reveals that the criterion for the Mott VRH conduction, $\overline{R}_{hop,Mott}/\xi > 1$, is satisfied for the samples Nos. 4–7 over the temperature intervals where the Eq. (1) is applicable. For the sample No. 3, however, the extracted value of $\overline{R}_{hop,Mott}/\xi$ is slightly too small to meet the Mott criterion. This may be understood in terms of this particular sample being just barely resistive enough for the Mott VRH conduction process to happen. Note that our evaluated values of $\overline{R}_{hop,Mott}$ in the samples Nos. 3–7 are much smaller than the film thicknesses, and hence our samples are three-dimensional with regard to the Mott VRH conduction process.

We now analyze the charge transport mechanism at the lower temperature region where notable deviations from the Mott VRH conduction law are evident. We recall that the Coulomb interactions between charge carriers were totally ignored and that $N_0(E_F)$ was treated as a constant in the Mott VRH conduction theory. If the long-range nature of the Coulomb interactions are taken into account, Efros and Shklovskii found that the electronic DOS in the vicinity of $E_F$ is no longer a constant, but is given by (in three dimensions)[10-12]

$$N(E) = \frac{3}{\pi}\left(\frac{\kappa}{e^2}\right)^3 (E - E_F)^2, \tag{5}$$

where $\kappa$ is the static dielectric constant. Therefore, the magnitude of $N(E = E_F)$ is zero, leading to a Coulomb gap with a width[10-12]

$$\Delta_{CG} = \frac{e^3 \sqrt{N_0(E_F)}}{\kappa^{3/2}}. \tag{6}$$

In this case, ES predicted that the temperature dependence of resistivity (in all dimensions) can be written as[10-12]

$$\rho_{ES} = \rho_{ES_0} e^{(T_{ES}/T)^{1/2}}, \tag{7}$$

where $\rho_{ES_0}$ is a resistivity parameter, and $T_{ES}$ is a characteristic temperature defined by[10-12]

$$T_{ES} = \frac{\beta_1 e^2}{\kappa \xi k_B}, \tag{8}$$

where $\beta_1$ is a constant with a value of $\simeq 2.8$. Combining Eqs. (2), (6) and (8), one obtains

$\Delta_{CG} \approx k_B (T_{ES}^3 / T_M)^{1/2}$. When the temperature $T$ is high enough for a hopping electron to explore the energy range $k_B (T^3 T_M)^{1/4} > \Delta_{CG}$, the influence of the Coulomb gap can be ignored and one recovers the Mott VRH conduction law described by Eq. (1). Below the crossover temperature, $T_{cross} = 16 T_{ES}^2 / T_M$, only the electronic states inside the Coulomb gap are accessible and the ES VRH conduction law, Eq. (7), is to be expected. Then, the most probable hopping distance and the average hopping energy are, respectively, given by[10-12]

$$\overline{R}_{hop,ES} = \frac{1}{4} \xi \left( \frac{T_{ES}}{T} \right)^{1/2}, \tag{9}$$

and

$$\overline{W}_{hop,ES} = \frac{1}{2} k_B T \left( \frac{T_{ES}}{T} \right)^{1/2}. \tag{10}$$

To check whether the ES VRH conduction process does happen in our high-resistivity ZnO films (Nos. 3–6) at low temperatures, we replot in Fig. 4 the variation of the logarithm of resistivity with $(1/T)^{1/2}$. Inspection of Fig. 4 clearly indicates that our measured $\log \rho$, or $\ln \rho$, varies linearly with $(1/T)^{1/2}$ at low temperatures. To determine the temperature ranges over which the ES VRH conduction process is valid in our films, we fitted the low temperature resistivity data of the samples Nos. 3–6 to Eq. (7). We found that the temperatures below which the Coulomb gap effect should be considered are approximately 5, 6, 10 and 20 K for the samples No. 3, No. 4, No. 5 and No. 6, respectively. (Owing to its relatively high resistivity, our lowest measurement temperature of 20 K for the sample No.7 was still not low enough to reveal a sufficiently wide temperature interval for a clear determination of the ES process. However, it is nature to expect that the Coulomb gap effect should already become important at temperatures ≥ 20 K.) Our fitted values for the relevant parameters are listed in Table II.

From the extracted values of $T_M$ and $T_{ES}$ for the Mott VRH and the ES VRH conduction laws, respectively, we have calculated the values of the crossover temperature $T_{cross}$ and the Coulomb gap $\Delta_{CG}$ in our films (see Table II). For the samples No. 3 and No.4, we found that our computed values of $T_{cross}$ fall within the temperature intervals where the hopping conduction behavior changes from the Mott law to the ES law. However, for the samples No. 5 and No. 6, our calculated values of $T_{cross}$ are higher than the minimum temperatures down to which the Mott VRH conduction law was experimental observed. This latter discrepancy can partly originate from the difficulties and the accompanied large uncertainties in extracting the values of $T_M$ and $T_{ES}$ by directly fitting experimental data to Eqs. (1) and (7). Such uncertainties could further be inherited from the fact that the change from the Mott VRH conduction behavior to the ES VRH conduction behavior is a smooth crossover, rather than a sharp transition. Therefore, any slight deviations of the exponents in Eq. (1) and Eq. (7) from 1/4 and 1/2, respectively, if happen, may cause the evaluated values of $T_{cross}$ and $\overline{R}_{hop,Mott}/\xi$ to be somewhat inconsistent with the criterions required for the two theories to be applicable.[16,17] We note that, from our fitted values of $T_{ES}$, we obtain the relation $\overline{W}_{hop,Mott} > \Delta_{CG} > \overline{W}_{hop,ES}$ in the films Nos. 3–6. This observation supports the assertion of a crossover from the Mott type to the ES type VRH conduction in our ZnO films. It should be pointed out that, although a crossover from the Mott to the ES VRH conduction processes has previously been observed in some typical semiconductor systems, such as CdSe[17] and Si:B,[18] the reported ES VRH conduction process only happened at very low temperatures (≤ 1 K). The high crossover temperatures in our ZnO films can be explained in terms of the high values of $N_0(E_F)$ associated with these samples, which in turn can be ascribed to originating from the presence of high donor concentrations due to oxygen deficiencies.

After the observations of a crossover from the Mott to the ES VRH conduction mechanisms in several semiconducting materials, Aharony, Zhang and Sarachik[19] proposed a phenomenological scaling relation to describe the overall temperature behavior of VRH resistivity from high to low temperatures as follows:

$$\ln(\rho/\rho_0) = A f(T/T_x), \quad (11)$$

where the scaling parameters $A$ and $T_x$ depend on the individual sample properties, but the function $f(x)$ is predicted to be universal and having the form

$$f(x) = \frac{1 + [(1+x)^{1/2} - 1]/x}{[(1+x)^{1/2} - 1]^{1/2}}. \quad (12)$$

We have fitted our measured resistivity data below 90 K for the samples Nos. 3–6 using Eqs. (11) and (12). Our results for each film are plotted in Fig. 5(a). In Fig. 5(b), we replot together the normalized resistivity versus normalized temperature for these four samples with double-logarithmic scales. The data collapse closely onto a single curve, as is evident in Fig. 5(b). Therefore, this scaling relation describes our results very well. In the comparison with the theoretical predictions, we first treated $A$, $T_x$ and $\rho_0$ as adjusting parameters in the least-squares fits. Then, we obtained the two new characteristic temperatures $T_M^{'}$ and $T_{ES}^{'}$ according to $T_M^{'} = A^4 T_x$ and $T_{ES}^{'} = 9 A^2 T_x / 2$ (Ref. 19). Here $T_M^{'}$ ($T_{ES}^{'}$) is nominally the Mott (ES) VRH conduction characteristic temperature. Table II lists our extracted values of $T_M^{'}$ and $T_{ES}^{'}$ for four samples. This Table reveals that our experimental values of $T_{ES}^{'}$ and $T_{ES}$ in every sample are comparable. However, our deduced value of $T_M^{'}$ is significantly lower than the corresponding value of $T_M$ for a given sample. Similar discrepancies between the values of $T_M^{'}$ and $T_M$ had also been noted in Ref. 19. An improved functional form for $f(x)$ derived by using a microscopic theory might resolve the discrepancies between the theory and experiment.[20] This issue requires further investigations.

Finally, in the samples No. 1 and No. 2, we found that neither the Mott nor the ES VRH conduction law can describe our measured temperature dependent resistivities. The resistivity increased only by a factor of 2.5 (4.6) between 300 K and 2 K in the sample No. 1 (No. 2).

Such small resistivity increases, together with the experimental values of carrier concentration $n$(300 K) listed in Table I, suggest that the two films lie very close to the metal-insulator transition, and hence the VRH conduction processes do not play a major role in determining the charge transport. [It is estimated that the metal-insulator transition in single-crystalline ZnO occurs at a critical carrier concentration $n_c \approx 5 \times 10^{18}$ cm$^{-3}$ (Refs. 21 and 22).] Moreover, since the individual ZnO crystallites in these two films must possess comparatively low resistivities due to their containing large amounts of oxygen deficiencies, the measured resistivities could have been largely weighed by the grain-boundary resistivities.[8] Such additional complications then make a quantitative analysis of the conduction processes in these two films less straightforward.

**IV. CONCLUSION**

We have systematically investigated the temperature dependent resistivities of polycrystalline ZnO films, which were deposited under different O$_2$ partial pressures, over a wide range of temperature from 300 K down to liquid-helium temperatures. Below about 100 K, the charge transport processes are found to be governed by the variable-range-hopping conduction mechanisms. In particular, two distinct temperature behaviors of resistivity, i.e., the Mott VRH hopping law $\ln \rho \propto (1/T)^{1/4}$ and the Efros-Shklovskii VRH hopping law $\ln \rho \propto (1/T)^{1/2}$ are sequentially observed as the temperature is decreased. To the best of our knowledge, this is the first experimental observation of the ES VRH conduction process, and of the presence of a Coulomb gap, in oxygen deficient ZnO films. Our experimental values of the crossover temperature between the Mott and the ES VRH conduction mechanisms are reasonably in line with the theoretical predictions. The temperature behavior of resistivity over the entire VRH conduction regime from the Mott type to the ES type processes can be closely described by a universal scaling relation.

**Acknowledgments**

This work was supported by the Taiwan National Science Council through Grant No. NSC 98-2120-M-009-004 and the MOE ATU Program (to JJL), and by the Key Project of Chinese Ministry of Education through Grant No. 109042 (to ZQL).

**Table I.** Values of relevant parameters for seven ZnO films studied in this work. The carrier concentrations $n$(300 K) were determined from the Hall effect measurements. The values of $\bar{W}_{hop,Mott}$ were calculated for a representative temperature of 40 K.

| Film No. | O$_2$ flux (SCCM) | thickness (nm) | $\rho$(300 K) ($\Omega$ cm) | $n$(300 K) (cm$^{-3}$) | $\rho_{M_0}$ ($\Omega$ cm) | $T_M$ (K) | $N_0(E_F)$ (J$^{-1}$ m$^{-3}$) | $\dfrac{\bar{R}_{hop,Mott}}{\xi}$ | $\bar{W}_{hop,Mott}$ (meV) |
|---|---|---|---|---|---|---|---|---|---|
| 1 | 0 | 1020 | 0.073 | $8\times10^{18}$ | — | — | — | — | — |
| 2 | 0.02 | 1060 | 0.476 | $6\times10^{18}$ | — | — | — | — | — |
| 3 | 0.10 | 1050 | 3.23 | $1\times10^{18}$ | 1.66 | 1250 | $1.3\times10^{47}$ | $2.23/T^{1/4}$ | 2.04 |
| 4 | 0.15 | 1090 | 11.4 | $6\times10^{17}$ | 3.53 | 5820 | $2.8\times10^{46}$ | $3.27/T^{1/4}$ | 2.99 |
| 5 | 0.25 | 1090 | 26.0 | $2\times10^{17}$ | 2.90 | 34500 | $4.7\times10^{45}$ | $5.11/T^{1/4}$ | 4.67 |
| 6 | 0.50 | 1030 | 48.4 | $2\times10^{16}$ | 1.82 | 108000 | $1.5\times10^{45}$ | $6.80/T^{1/4}$ | 6.21 |
| 7 | 0.80 | 951 | 206 | $3\times10^{15}$ | 4.18 | 264000 | $6.2\times10^{44}$ | $8.50/T^{1/4}$ | 7.77 |

**Table II**. Values of relevant parameters for four ZnO films in which the Efros-Shklovskii variable-range-hopping conduction process is observed. The values of $\bar{W}_{hop,ES}$ were calculated for a representative temperature of 5 K.

| Film No | $\rho_{ES}$ ($\Omega$ cm) | $T_{ES}$ (K) | $\Delta_{CG}$ (meV) | $\dfrac{\bar{R}_{hop,ES}}{\xi}$ | $\bar{W}_{hop,ES}$ (meV) | $T_{cross}$ (K) | $T'_M$ (K) | $T'_{ES}$ (K) |
|---|---|---|---|---|---|---|---|---|
| 3 | 8.45 | 36.1 | 0.60 | $1.50/T^{1/2}$ | 0.60 | 16.6 | 533 | 40.7 |
| 4 | 25.3 | 113 | 1.36 | $2.66/T^{1/2}$ | 1.03 | 35.2 | 1330 | 130 |
| 5 | 25.3 | 387 | 3.34 | $4.92/T^{1/2}$ | 1.89 | 69.4 | 711 | 385 |
| 6 | 53.9 | 601 | 3.87 | $6.13/T^{1/2}$ | 2.37 | 53.5 | 3100 | 684 |

**Figure Captions**

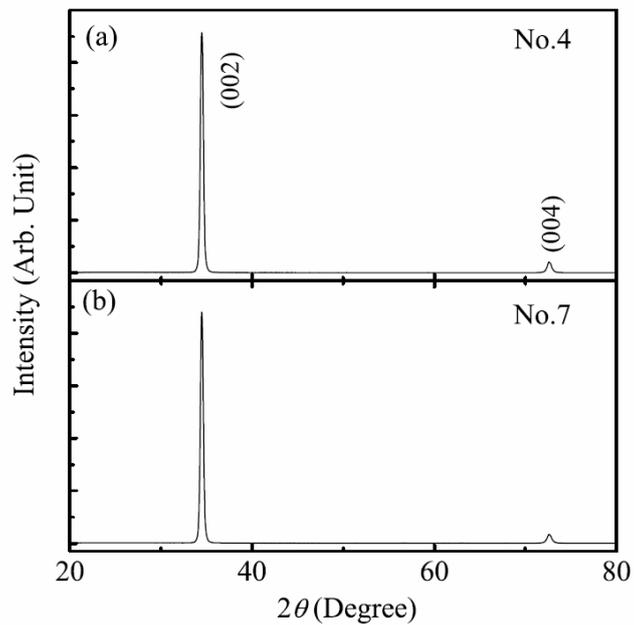

**Figure 1.** XRD patterns for two representative ZnO films, as indicated.

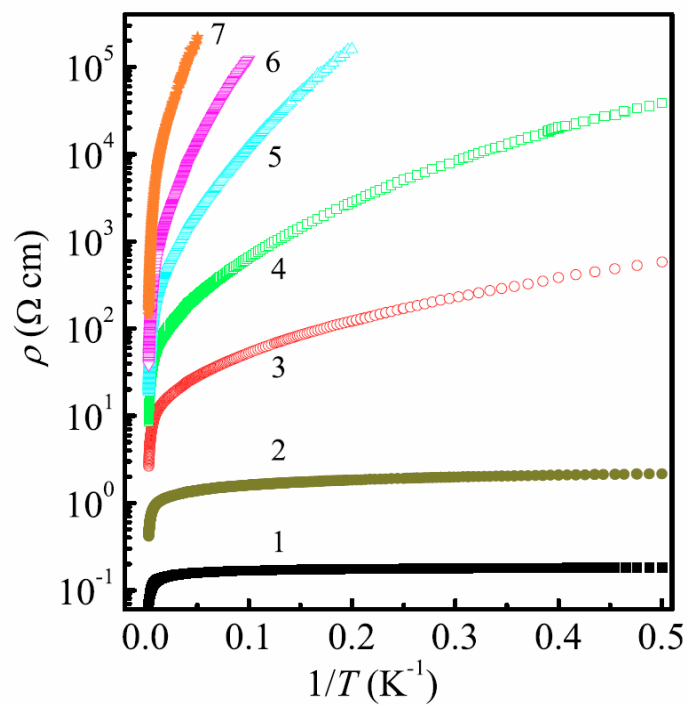

**Figure 2.** Variation of the logarithm of resistivity with reciprocal temperature for seven ZnO films, as indicated.

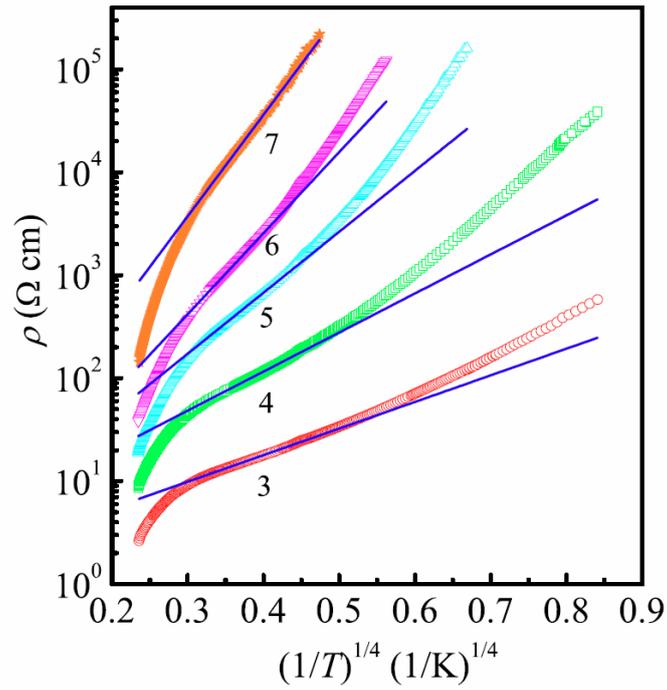

**Figure 3.** Variation of the logarithm of resistivity with $(1/T)^{1/4}$ for five ZnO films, as indicated. The straight solid lines are the least-squares fits with the Mott VRH conduction law, Eq. (1), with the values of the fitting parameters listed in Table I.

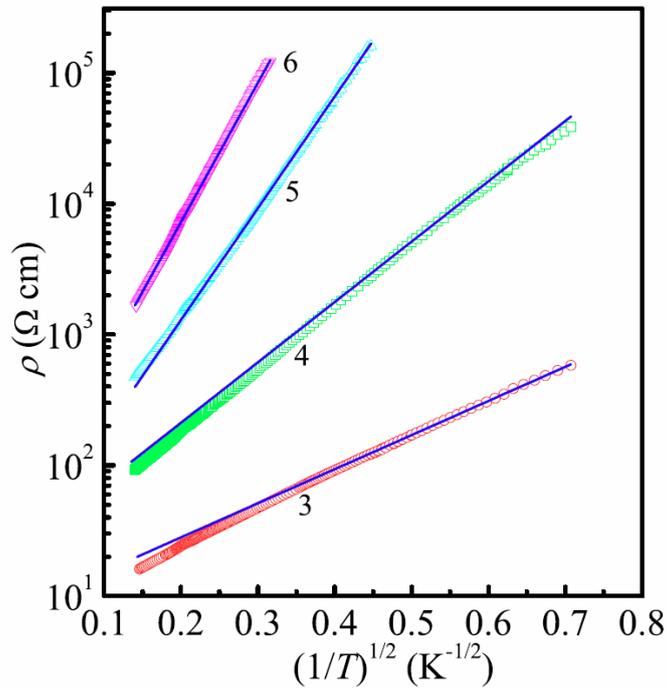

**Figure 4.** Variation of the logarithm of resistivity with $(1/T)^{1/2}$ for four ZnO films, as indicated. The straight solid lines are the least-squares fits with the Efros-Shklovskii VRH conduction law, Eq. (7), with the values of the fitting parameters listed in Table II.

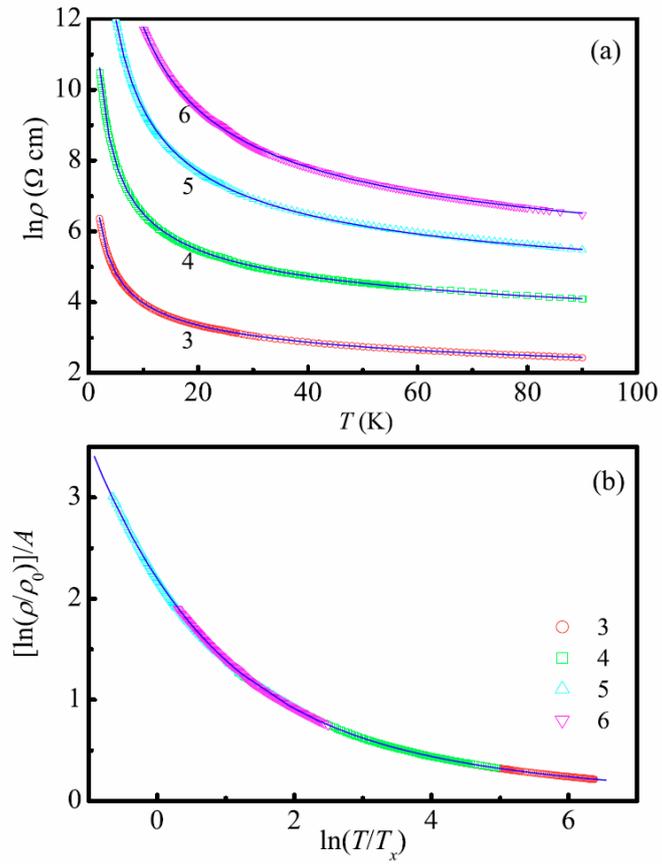

**Figure 5.** (a) Variation of $\ln\rho$ with temperature for four ZnO films, as indicated. (b) Double-logarithm plot of normalized resistivity versus normalized temperature for the same four films. Notice that the data collapse closely. In (a) and (b), the symbols are the experimental data, and the solid curves are least-squares fits to the theoretical predictions of Eqs. (11) and (12).